\documentstyle[aps,preprint,prd,epsf]{revtex}

\tighten
\begin{document}

\preprint{TUM--T31--43/93}
\title{The Decay $B\to K^*\gamma$ from QCD Sum Rules}
\author{Patricia Ball}
\address{Physik--Department, TU M\"unchen, D--85747 Garching, FRG}
\date{August 7, 1993}
\maketitle
\begin{abstract}
Using the QCD sum rules approach and the complete leading--logarithmic
short--distance coefficients determining the penguin--amplitude, we
obtain $B(B\to K^*\gamma) = (6.8\pm 2.4)\cdot 10^{-5}$ for top--quark
masses in the range (150--200) GeV and $|V_{cb}| = 0.035$ for $\tau_B =
1.5\,\text{ps}$.  The ratio to the inclusive rate is
$B(B\to K^*\gamma)/B(B\to X_s\gamma) = (20\pm 6)\% $.
\end{abstract}
\vspace{1cm}

Since some time, the penguin--induced decay $b\to s\gamma$ has
received continuous interest as a possible probe of new physics
(cf.\ e.g.\ \cite{grinstein,BG93,techno}). The interest quite 
naturally increased when experimental results on the exclusive 
radiative decay $B\to K^* \gamma$ became available \cite{Exp}. 
The interpretation of these results, however, is not obvious, 
since existing calculations focused mainly on the inclusive 
decay $B\to X_s\gamma$ with inidentified strange hadrons $X_s$ in 
the final state. The rate of this decay can  be obtained from the 
perturbatively calculable short--distance expansion of the effective 
Hamiltonian. The existing estimates of hadronic corrections rely 
mainly on quark model \cite{alto,desh,AG91} as well as
QCD sum rule calculations \cite{dompfaff,AOS90}, and yield
rather different results. To be specific, theory is asked to provide 
one with the hadronic matrix element $\langle\, K^*(p_{K^*})\,|\,\bar
s\sigma_{\mu\nu}q^\mu R b\,|\,B(p_B)\,\rangle$ at vanishing momentum
transfer squared, $q^2=(p_B-p_{K^*})^2=0$, where $R=(1+\gamma_5)/2$ 
is the usual projector on right--handed quarks, and $\sigma_{\mu\nu} = 
i/2\,[ \gamma_\mu, \gamma_\nu]_-$. In this letter, we aim at to give 
an updated analysis of the exclusive decay $B\to K^* \gamma$, using 
both short--distance coefficients in leading--order accuracy 
\cite{M93} and improved QCD sum rules.\footnote{After finishing our 
calculations, we became aware of Ref.\ \cite{martinelli} where it is 
claimed that the Wilson coefficients calculated in \cite{M93} are not 
completely correct. The change in numerics is, however, negligible 
($\sim 1$\%).} We will not explain the  QCD sum rules approach in 
detail, but just mention that it has established itself as a reliable 
tool to infer the gross features of QCD induced non--perturbative 
dynamics of hadronic matrix elements. The method relies on the
field--theoretical aspects and features of QCD and was designed to
make maximum use of known manifestations of non--perturbative QCD. 
Originally invented for the calculation of vacuum--to--meson
transition amplitudes \cite{SVZ}, it soon found application to the 
calculation of the electromagnetic form factor of the pion 
\cite{smilga} and other meson--to--meson transition amplitudes (cf.\ 
\cite{shifman} for a review). Although QCD sum rules in general 
yield less detailed results than fine--tuned models, they have 
the advantage that only a small number of parameters is needed that 
have an evident physical meaning (e.g.\ quark masses) and/or 
characterize the non--perturbative regime of QCD (e.g.\ the 
so--called quark condensate, the order parameter of chiral symmetry 
breaking). Once these parameters are fixed from well known processes,
they can be used to calculate for instance heavy meson decays.

The decay we are interested in is determined by the effective
Hamiltonian 
\begin{equation}
H_{\text{eff}} = \frac{4G_F}{\sqrt{2}}\,V_{cb}V_{cs}^*
\sum\limits_{i=1}^8 C_i(\mu) {\cal O}_i(\mu),
\end{equation}
describing an effective five--quark theory where the effects of the 
top--quark and the W--boson are integrated out. The operators 
${\cal O}_i$ mix under renormalization, their anomalous dimensions are
known completely to leading--order and partly to next--to--leading 
order accuracy \cite{grinstein,M93,M91}. Since a calculation of 
$O(\alpha_s)$ corrections to our sum rules is beyond the scope of this
letter, we for consistency only use the leading--order coefficients 
calculated in \cite{M93}. 

The only operator relevant for the decay $B\to K^* \gamma$ is
\begin{equation}
{\cal O}_7 = \frac{e}{16\pi^2}\,\bar s \sigma_{\mu\nu} (m_bR +
m_s L) b F^{\mu\nu}.
\end{equation}
Here $m_b$ and $m_s$ are the masses of the b- and s-quark field,
respectively, $F_{\mu\nu}$ is the electromagnetic field strength
tensor, $L=(1-\gamma_5)/2$ the projector on left--handed quarks.
In calculating the decay rate, we note that the structure
$\sigma_{\mu\nu}\gamma_5$ is not independent of $\sigma_{\mu\nu}$, but
can be expressed as $\sigma_{\mu\nu}\gamma_5=-i
\epsilon_{\mu\nu\alpha\beta}\sigma^{\alpha\beta}/2$, and thus the form
factor decomposition of $\langle\,K^*,\lambda\,|\,\bar s
\sigma_{\mu\nu}q^\mu b\,|\,B\,\rangle$ determines the decay rate
completely. With 
\begin{equation}\label{eq:ME}
\langle\,K^*,\lambda\,|\,\bar s \sigma_{\mu\nu}q^\mu b\,|\,B\,
\rangle= -i F(q^2,\mu)\,\epsilon_{\alpha\nu\gamma\delta}
\epsilon_{K^*}^{(\lambda)*\alpha}p_B^\gamma p_{K^*}^\delta,
\end{equation}
where $q=p_B-p_{K*}$ is the photon momentum, $\lambda$ denotes the 
helicity state of the K$^*$ and $\epsilon^{(\lambda)}_\mu$ is the 
corresponding polarization vector, we find
\begin{equation}\label{eq:rate}
\Gamma(B\to K^* \gamma) = \frac{\alpha}{128\pi^4}\,G_F^2 \,|V_{cs}^*
V_{cb}|^2 \,C_7^2(\mu) F^2(0,\mu) (m_b^2+m_s^2)\,
\frac{(m_B^2-m_{K^*})^3}{m_B^3}\,.
\end{equation}
In (\ref{eq:ME}) and (\ref{eq:rate}), we have made explicit the 
scale--dependence of the matrix element by
ascribing the form factor a dependence on the normalization scale $\mu$.
The inclusive rate is given by
\begin{equation}\label{eq:rateinklu}
\Gamma(B\to X_s\gamma) = \frac{\alpha}{32\pi^4}\,G_F^2 \,|V_{cs}^*
V_{cb}|^2 \,C_7^2(\mu)\,m_b^2\,m_B^3,
\end{equation}
neglecting the s--quark mass.

In the framework of QCD sum rules, the form factor $F(0,\mu)$
 can be calculated as ground state contribution to the
three--point correlation function
\begin{eqnarray}
\lefteqn{T_{\alpha\nu}(p_B^2,p_{K^*}^2,q^2=0) =}\nonumber\\
& = &  \left.i^2\!\! \int\!\! d^4\!x\, d^4\!y\, 
e^{-ip_Bx+ip_{K^*}y}\,
\langle\, 0 \, |\, T\, \bar q(y) \gamma_\alpha s(y)\,\bar s(0) q^\mu
\sigma_{\mu\nu} b(0)\,\bar b(x) i\gamma_5
q(x)\,|\,0\,\rangle\right|_{q^2=0}\nonumber\\
& = & iT(p_B^2,p_{K^*}^2) \epsilon_{\alpha\nu\beta\gamma}
p_B^\beta p_{K^*}^\gamma.\label{eq:corr}
\end{eqnarray}
$T$ can be expressed as (double) dispersion relation where the double
spectral function is given by the contribution of intermediate
on--shell particles coupling to the currents. In zero--width
approximation we find
\begin{equation}
T_{\alpha\nu} = -i\,F(0,\mu)\, \sum\limits_\lambda \frac{\langle\,0\,|
\,\bar q \gamma_\alpha s\,|\,K^*,\lambda\,\rangle\,\epsilon_{\alpha
\nu\gamma\delta}
\epsilon_{K^*}^{(\lambda)*\alpha}p_B^\gamma p_{K^*}^\delta
\langle\,B\,|\,\bar b i\gamma_5 q\,|\,0\,\rangle}{(p_B^2-m_B^2)
(p_{K^*}^2-m_{K^*}^2)} + \text{continuum},
\end{equation}
where the first term on the right--hand side just contains the
expression we are interested in, the second one stands for the
contributions of higher resonances and many--particle states.
Introducing the meson leptonic decay constants $f_B$ and $f_{K^*}$ for
the B and the K$^*$ meson, respectively, as
\begin{equation}
\langle\, 0\,|\, \bar qi\gamma_5b\,| \,B\,\rangle = \frac{m_B^2
f_B}{m_b},\qquad \langle\, 0\,|\, \bar q \gamma_\mu s\,| \,
K^*,\lambda\,\rangle = m_{K^*} f_{K^*} \epsilon_\mu^{(\lambda)},
\end{equation}
we find
\begin{equation}\label{eq:SRroh}
T = F(0,\mu)\, \frac{m_{K^*} f_{K^*} m_B^2 f_B}{m_b}\,
\frac{1}{(p_B^2-m_B^2)(p_{K^*}^2-m_{K^*}^2)} + \text{continuum}.
\end{equation}
Using quark--hadron duality, we model the continuum part by the 
purely perturbative double spectral
function above certain thresholds in the dispersion variables, the
so--called continuum thresholds, $s_B^0$ and $s_{K^*}^0$,
respectively. In order to enhance the contribution of the 
ground state and to suppress the continuum contribution, we subject 
$T$ to a Borel transformation $\hat B$ (cf.\ \cite{SVZ}) yielding
\begin{equation}
\hat B\,\frac{1}{s-p^2} = \frac{1}{M^2}\,e^{-s/M^2}.
\end{equation}
By means of this transformation, the variable $p^2$ gets effectively
replaced by $M^2$, the Borel parameter. Evidently, large
values of the dispersion variable $s$, i.e. the continuum
contribution, get exponentially suppressed.

Whereas $T$ is calculable within pure perturbation theory for
$p_B^2,\,p_{K^*}^2\to -\infty$ only, the QCD sum rules approach allows
one to get
closer to the more interesting physical region $p_B^2-m_b^2 \approx
p_{K^*}^2-m_s^2 \approx 0$. Performing an operator product expansion of
(\ref{eq:corr}), it proves feasible to parametrize the unknown 
long--distance behaviour by vacuum expectation values of certain 
gauge--invariant operators, the so--called condensates \cite{SVZ}. 
These matrix elements vanish, when taken over the perturbative vacuum,
but acquire finite values over the physical vacuum. Thus $T$ can be 
expressed as
\begin{equation}
T(p_B^2,p_{K^*}^2) = \sum_n T^{(n)}(p_B^2,p_{K^*}^2) \langle\, 
{\cal O}_n\,\rangle.
\end{equation}
The coefficient functions $T^{(n)}$ contain the short--distance
behaviour of the correlation function and are calculable
perturbatively. In this letter we take into account all operators
up to dimension n=6, i.e.\ the quark, gluon, mixed, and four--quark
condensates. The evaluation of these contributions proceeds along 
standard
lines, for the coefficient of the gluon condensate we use the method
proposed in \cite{ball}. All coefficients are calculated to
$O(\alpha_s^0)$ and including their full dependence on $m_s$. For
lack of space, we do not give the formul\ae\/ in this letter.

Applying the Borel transformation in both the variables $p_B^2$ and
$p_{K^*}^2$, (\ref{eq:SRroh}) yields the final sum rule
\begin{equation}\label{eq:SR}
F(0,\mu) = \frac{m_b}{m_{K^*}f_{K^*}m_B^2f_B}\,e^{m_B^2/M_B^2+
m_{K^*}^2/M_{K^*}^2}\,M_B^2 M_{K^*}^2\,\hat T(M_B^2,M_{K^*}^2).
\end{equation}
Here $\hat T$ is the correlation function with subtracted continuum
contribution, $M_B^2$ and $M_{K^*}$ are the Borel parameters. At this
point, a few remarks are in order. First, the quark mass $m_b$
appearing in Eq.~(\ref{eq:SR}) is the renormalization scheme and group
invariant pole mass, related to the running mass in the
$\overline{\text{MS}}$ scheme by
\begin{equation}
m_{\text{pol}} = m_{\overline{\text{MS}}}(\mu)\left( 1+
\frac{\alpha_s(\mu)}{\pi}\,\left \{ \frac{4}{3} + 
\ln \frac{\mu^2}{m^2_{\overline{\text{MS}}}} \right\}\right).
\end{equation}
In the numerical evaluation we use the value $m_b = 4.8\,\text{GeV}$.
For the mass of the s--quark, we use the running $\overline{\text{MS}}$
mass with $m_s(\mu=1\,\text{GeV})=(0.20\pm0.05)\,\text{GeV}$.
Actually, the sum rule proves not very sensitive to that
value. For the condensates, we use the standard set of values (at the
renormalization point $\mu = 1\,\text{GeV}$):
\begin{eqnarray}
\langle\,\bar q q\,\rangle =  (-0.24\,\text{GeV})^3\,, & \qquad &
\langle\, \bar q g \sigma_{\mu\nu}G^{\mu\nu} q\,\rangle =
0.8\,\text{GeV}^2 \,\langle\,\bar q q\,\rangle,\nonumber\\
\langle\, \alpha_s/\pi^2\,G^2\,\rangle = 0.012\,\text{GeV}^4\,, & \qquad
& \pi\alpha_s \langle \bar q
      \gamma^\tau \lambda^A q \sum_{u,d,s}
      \bar q \gamma_\tau \lambda^A q \rangle \approx -\frac{16}{9}\, \pi
\alpha_s \langle \bar q q \rangle^2,\nonumber\\
4\pi\alpha_s \langle \bar d \bar u u d \rangle \approx 4\pi
\alpha_s
\langle \bar q q \rangle^2.& &
\end{eqnarray}
For $f_{K^*}$, we use the experimental value extracted from $\tau^-\to
K^{*-} \nu_{\tau}$, $f_{K^*} = 0.21\,\text{GeV}$  \cite{PDG}. For
$f_B$, we use a two--point sum rule (see \cite{alel}, e.g.) without
radiative corrections. In the limit of infinitely heavy quarks, this
procedure is required to obtain the right normalization of the form
factor \cite{radi}, and largely reduces the dependence of the sum rule
on the Borel parameter and the influence of (in our case unknown)
radiative corrections \cite{bagan}. We take over the procedure to the
present case whith one light quark, and use half the value of the
Borel parameter $M_B^2$ in the sum rule for $f_B$ and equal continuum
thresholds $s_B^0$ in both the two-- and the three--point sum rule. In
addition, we evaluate (\ref{eq:SR}) at a fixed ratio $M_B^2/M_{K^*}^2
= 3$. The numerical values of the continuum thresholds are taken from
the region of maximum stability in the two--point sum rules, where one
finds $s_{K^*}^0 = 1.7\,\text{GeV}^2$ and $s_B^0 = 34\,\text{GeV}^2$,
and a ``sum rule window'' $7\,\text{GeV}^2 \lesssim M_B^2 \lesssim
10\, \text{GeV}^2$ (cf.\ \cite{BBD,BBBD}).

Next, we have to say some words about the residual scale--dependence
of $F(0,\mu)$. In contrast to calculations at partonic level, the 
condensates induce a scale--dependence of the sum rule already at
leading order in $\alpha_s$. Whereas previous sum rule calculations 
tacitly assumed $\mu=m_b$ \cite{dompfaff,AOS90}, we argue that the 
effective scale is much below that. This becomes plausible when one 
takes into account that the sum rule calculation is an intrinsic 
off--shell one where the correlation function is calculated 
in the not so deep Euclidean region of external momenta and then 
analytically continued to the physical region. Thus the characteristic
scale is rather given by the virtualities of the particle 
currents than by a singularity of the correlation function at
threshold. Our reasoning gets support from the analysis of
the heavy quark limit where the use of a low normalization
scale $M^2/(2m_Q)\approx 1\,\text{GeV}$ in the radiative corrections 
to the sum rule for $f_B$ is absolutely stringent \cite{BBBD}. From 
all that we conclude that the proper
normalization scale of the sum rule is coupled to the Borel
parameters. In view of the successes of previous form factor
determinations (\cite{BBD}, e.g.), we conform to the choice
made there and use $\mu = (M_B^2 M_{K^*}^2)^{1/4}$ which typically is
about $2\,\text{GeV}$. Note that a further discussion
 of that point would require the complete calculation of
$O(\alpha_s)$ corrections to the correlation function which is
complicated by the effects of operator mixing. 

\begin{figure}[h]
\centerline{
\epsfbox{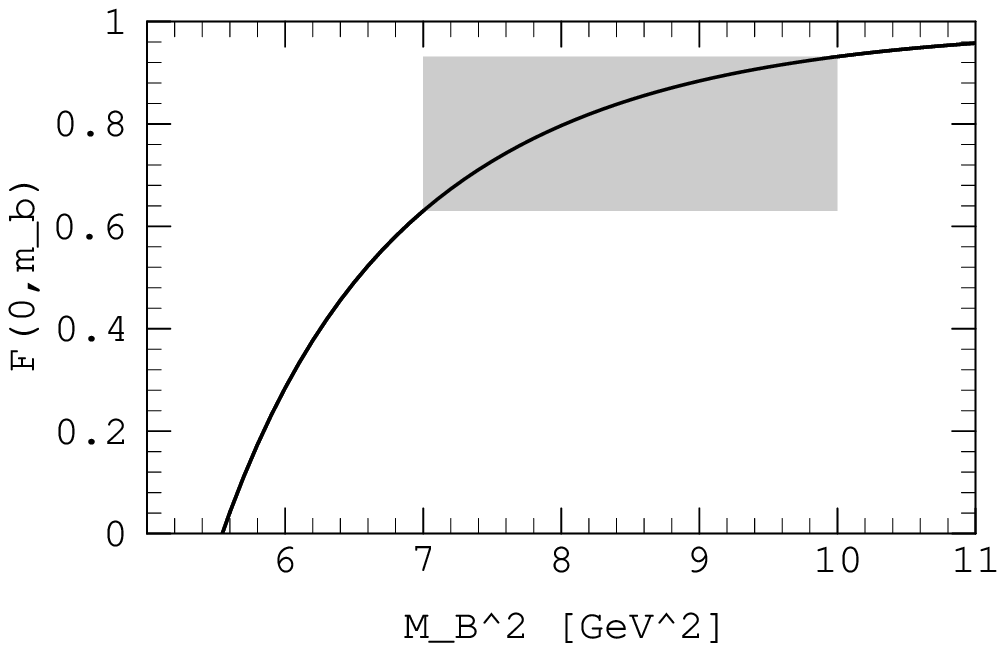}}
\vspace{-0.4in}
\caption[]{The form factor $F(0,m_b)$ as function of the Borel
parameter $M_B^2$ for $m_b = 4.8\,\text{GeV}$, $m_s(1\,\text{GeV}) =
0.2\,\text{GeV}$ and continuum thresholds $s_B^0 = 34\,\text{GeV}^2$,
$s_{K^*}^0 = 1.7\,\text{GeV}^2$, evaluated at a fixed ratio of the
Borel parameters, $M_B^2/M_{K^*}^2 = 3$. The shaded area indicates the
{}``sum rule window''.}\label{fig:1}
\vspace{-0.4in}
\centerline{
\epsfbox{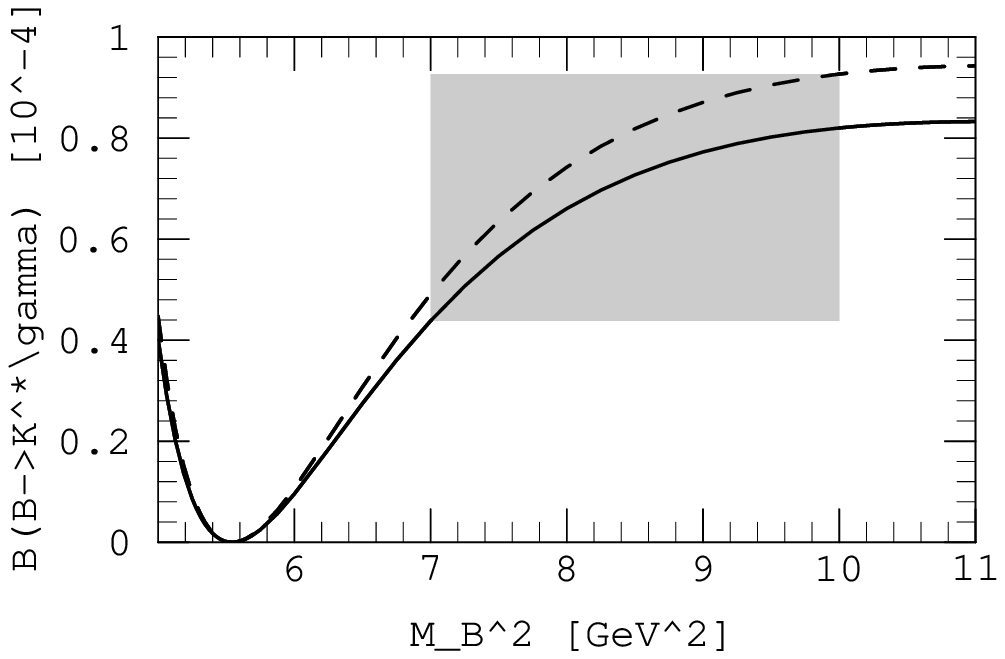}}
\vspace{-0.4in}
\caption[]{The branching ratio $B(B\to K^*\gamma)$ as function of the
Borel parameter $M_B^2$ with the same parameters as in
Fig.~\protect{\ref{fig:1}}. The shaded area again indicates the
``sum rule window''. The solid line corresponds to $m_{\text{top}} =
150\,\text{GeV}$, the dashed line to $m_{\text{top}}=200\,\text{GeV}$.
}\label{fig:2}
\end{figure}

The next step now is to disentangle the intrinsic $M^2$--dependence of
the sum rule for the form factor reflecting the inherent uncertainty
of the QCD sum rule method from the scale--dependence induced by 
renormalization group. This amounts to scaling up $F(0,\mu)$ from the 
lower scale $\mu$ to, say, $m_b$. In doing so, we note that the complete
solution of the renormalization group equation for ${\cal O}_7$,
\begin{equation}
{\cal O}_7(m_b) = {\cal O}_7(\mu)\left[
\frac{\alpha_s(m_b)}{\alpha_s(\mu)} \right]^{16/(3\beta_0)} +
\sum\limits_{i\neq 7} \epsilon_i(\mu) {\cal O}_i(\mu)
\end{equation}
with coefficients $\epsilon_i(\mu)$ that describe the admixture of the
other operators at $O(\alpha_s)$,
simplifies drastically for the matrix elements, since 
\begin{equation}\label{eq:ME2}
\langle\, K^*\gamma\,|\,{\cal O}_i\,|\,B\,\rangle = 0\quad
\text{for\,} i\neq 7
\end{equation}
at the specific scale $\mu$ and at one--loop level. Thus we have
\begin{equation}\label{eq:FF}
F(0,m_b) = F(0,\mu)\left[ \frac{\alpha_s(m_b)}{\alpha_s(\mu)} 
\right]^{16/(3\beta_0)}.
\end{equation}
Note, however, that although the initial condition for the solution
of the renormalization group equation contains seven zeros (the matrix
elements in (\ref{eq:ME2})), all these matrix elements acquire
non--zero values in scaling up to $m_b$ and contribute to the rate. In
Fig.~\ref{fig:1} we show $F(0,m_b)$, calculated according to
Eq.~(\ref{eq:SR}) and Eq.~(\ref{eq:FF}) as function of the Borel 
parameter $M_B^2$. The shaded area shows the working region of the sum
rule, $7\,\text{GeV}^2\leq M_B^2 \leq 10\,\text{GeV}^2$.
Unfortunately, the sum rule is rather unstable against variation of
the Borel parameter which is due to a relative sign between the
contributions of the quark and the mixed condensate. This causes these
contributions to be the dominant ones, whereas the perturbative
contribution is rather small ($\sim 20\%$), the contribution of the
gluon condensate and the four quark condensates are negligible ($\sim
2\%$). From the figure, we get
\begin{equation}
F(0,m_b) = 0.75\pm 0.10.
\end{equation}
This value is smaller than the one obtained in previous analyses
within the QCD sum rules approach \cite{dompfaff,AOS90}, and larger
than the value obtained in \cite{alto,desh}.

As mentioned before, it would be inconsistent to use this value for
the determination of the decay rate. The correct procedure is to
insert $F(0,\mu)$ as obtained from (\ref{eq:SR}) into
Eq.~(\ref{eq:rate}), using the same scale in the coefficient 
$C_7(\mu)$. 
For the combination $|V_{cb}|^2\,\tau_B$ that enters the formula for
the branching ratio, we use the value $1.81\cdot 10^{-15}s$ obtained
in \cite{ballvcb} which amounts to $|V_{cb}| = 0.035$ for the most
recent determination $\tau_B=1.5\,\text{ps}$ of the B meson lifetime
\cite{life}. For the mass of the t-quark that dominates the
penguin--amplitude we use the range of values $(150-200)\,\text{GeV}$
which is consistent with a recent determination of a lower bound on
$m_{\text{top}}$ from low--energy data \cite{Buras}.
Taking all together, Eq.~(\ref{eq:rate}) yields the branching ratio
$B(B\to K^*\gamma)$ plotted in Fig.~\ref{fig:2} as a function
of the Borel parameter $M_B^2$. Again we observe a rather strong
dependence on $M_B^2$ which exceeds all other
uncertainties in the input parameters of the sum rule, i.e.\ the quark
masses and the values of the quark and the mixed condensate. Thus we
get a branching ratio with large errors:
\begin{equation}
B(B\to K^*\gamma) = (6.8\pm 2.4)\cdot 10^{-5}
\end{equation}
which coincides within the errors with the experimental value $B(B\to
K^*\gamma) = (4.5\pm 1.5\pm 0.9)$. Taking the ratio of the exclusive 
to the inclusive decay rate, we obtain from Eqs.~(\ref{eq:rate}) 
and~(\ref{eq:rateinklu}):
\begin{equation}
\frac{B(B\to K^*\gamma)}{B(B\to X_s\gamma)} = (20\pm 6)\%.
\end{equation}
This value again is lower than the QCD sum rule results
\cite{dompfaff,AOS90} where values between 28\% and 39\% are given,
but it is much larger than old constituent quark model results (4.5\%
\cite{alto}, 6\% \cite{desh}). It agrees with a recent analysis in
\cite{AG93} where $(13\pm 3)\%$ is obtained.

Concluding, we remark that although our results are by far not
accurate enough as to impose constraints on $m_{\text{top}}$ or
possible new physics, we find satisfying agreement of the QCD sum rule
calculation for $B\to K^*\gamma$ with experiment. The main uncertainty
of the sum rule calculation is caused by different signs in two major
contributions which spoils the stability of the sum rule in the Borel
parameter. Although this uncertainty might be reduced by including
radiative corrections, their calculation proves prohibitively 
complicated.

\bigskip

\noindent {\bf Acknowledgement}: The author thanks M. Misiak for useful
discussions about the short--distance expansion for $b\to s\gamma$.

\end{document}